\documentclass[%
 reprint,
nofootinbib,
 amsmath,amssymb,
 aps,
]{revtex4-2}
\usepackage{amsmath,amsfonts,color}
\usepackage{xcolor,soul,amssymb,amsthm,amsmath,dsfont} 
\usepackage{tikz} 
\usepackage{color}
\usepackage{graphicx,algorithm,algorithmic}
\graphicspath{{figures/}{./}}
\usepackage{array}
\usepackage{eqparbox}
\usepackage{url}
\usepackage{relsize}
\usepackage{stfloats}
\usepackage{circuitikz}
\usepackage{graphicx}
\usepackage{dcolumn}
\usepackage{bm}
\usepackage{xcolor,soul,amssymb,amsthm,amsmath,dsfont} 
\usepackage{tikz} 
\usepackage{color}
\usepackage{graphicx,algorithm,algorithmic}
\graphicspath{{figures/}{./}}
\usepackage{array}
\usepackage{eqparbox}
\usepackage{url}
\usepackage{relsize}
\usepackage{comment}
\usepackage[normalem]{ulem}
\usepackage{soul}
\usepackage{mathtools,lipsum}

\newcommand{\x}{{\mathbf x}}
\newcommand{\y}{{\mathbf y}}
\newcommand{\black}{\color{black}}

\begin{document}
\preprint{APS/123-QED}

\title{Wasserstein speed limits for Langevin systems}
\author{Ralph Sabbagh, Olga Movilla Miangolarra, and Tryphon T. Georgiou}
\affiliation{Department of Mechanical and Aerospace Engineering, University of California, Irvine, California 92697, USA}

\begin{abstract}
Physical systems transition between states with finite speed that is limited by energetic costs. In this {\black work}, we derive bounds on transition times for general Langevin systems that admit a decomposition into reversible and irreversible dynamics, in terms of the Wasserstein distance between states and the energetic costs associated with respective reversible and irreversible currents. For illustration we discuss Brownian particles subject to arbitrary forcing \textcolor{black}{and an RLC circuit with time-varying inductor.}
\end{abstract}

\maketitle

\section{Introduction}

Over the latter part of the twentieth century, the quest to quantify energetic costs and the timing of thermodynamic transitions far from equilibrium, has given rise to Stochastic Thermodynamics \cite{sekimoto2010stochastic,seifert2012stochastic}. 
This discipline lies at the confluence of statistical mechanics and stochastic control, and has enabled effective models and exact results that pertain to microscopic thermodynamic ensembles subject to thermal fluctuations.

A most insightful discovery has been to express the minimal dissipation in thermodynamic transitions of overdamped particles as a path length traversed in a suitably metrized space of thermodynamic states \cite{aurell2011optimal,aurell2012refined}. Thus, for overdamped dynamics, entropy production serves as the long-sought notion of thermodynamic length \cite{weinhold1975metric,Ruppeiner1995geom,Crooks2007length,CrooksLengthControl2012} and has received considerable attention~\cite{dechant2019thermodynamic,Bradner2020geom,frim2021geometric,dechant2021geometric,taghvaei2021relation,EnergyHarvestingAnisotropic2021,dechant2022minimum,W2NakazatoEntropy2021,MarkovianW2Hasegawa2021,abiuso2022W2Carnotthermodynamics}. 

The enabling new insight took the form of a bound,
\begin{subequations}
\begin{align}
    \Sigma \geq \cfrac{\gamma}{\tau T}\mathcal{W}_2(\rho_0,\rho_\tau)^2,\label{OD}
\end{align}
 on the total entropy $\Sigma$ produced
 when steering a collection of overdamped particles via position-dependent forces, in a heat bath of temperature $T$ {\black and friction coefficient $\gamma$. The} 
 distance $\mathcal{W}_2(\rho_0,\rho_\tau)$ between end-point thermodynamic states $\rho_0$ and $\rho_\tau$ is known as the Wasserstein metric~\cite{villani2003topics}. In turn,
 \eqref{OD} readily provides a lower bound on the time $\tau$ of transition {\black in terms of the total entropy produced $\Sigma$}, as well as, a tight expression of the Wasserstein speed \cite{PhysRevResearch.3.043093} at time $t$, 
\begin{align}
\lim_{\tau\rightarrow 0^+}\frac{\mathcal{W}_2(\rho_t,\rho_{t+\tau})}{\tau}\leq \sqrt{\cfrac{T}{\gamma}\; \sigma_t} ,\label{ODb}
\end{align}
\end{subequations}
{\black where $\sigma_t$ is} the entropy production rate.
{\color{black}
Thermodynamic speed limits such as this one have had several applications, including the refinement of Landauer's thermodynamic cost of bit erasure \cite{aurell2011optimal,proesmans2020optimal,van2022finite} and estimation of relevant thermodynamic quantities, such as free energy and dissipation, in experimental settings \cite{gore2003bias,li2019quantifying,van2020entropy}.
}

The salient ingredient in \eqref{OD} is that
the Wasserstein metric is the {\em minimal} of an action integral~\cite{Benamou2000ACF} involving the probability current $\mathbf{J}$ that drives the thermodynamic state from $\rho_0$ to $\rho_\tau$ via the Fokker-Planck dynamics
$$\partial_t\rho = -\nabla\cdot\mathbf{J}.$$In the overdamped regime, that same action integral captures entropy production, in that
\begin{equation}\label{eq:Sigma}
    \Sigma = \cfrac{\gamma}{T}\int_0^{\tau}\int \cfrac{\|\mathbf{J}\|^2}{\rho}~\text{d}x\text{d}t.
\end{equation}
The term \textit{irreversible} is typically ascribed to
probability currents that satisfy such a relation.

The simple relation \eqref{eq:Sigma} is no longer valid in the underdamped setting where only a portion of the probability current in the phase plane is directly responsible for entropy production.  This is formalized for a general class of Langevin dynamics by defining even and odd degrees of freedom~\cite{PhysRevE.85.051113}, such as position and velocity, that leads to a decomposition of the probability current
$$
\mathbf{J} = \mathbf{J}^{\text{irr}} + \mathbf{J}^{\text{rev}}
$$
into
\textit{irreversible} and \textit{reversible} components, so that 
\begin{align*}
    \partial_t\rho = -\nabla\cdot\mathbf{J},\mbox{ while }\Sigma =\int_{0}^{\tau}\int \cfrac{\|\mathbf{J}^{\text{irr}}\|_{\mathbf{M}}^2}{\rho}~\text{d}x\text{d}v\text{d}t,
\end{align*}
\textcolor{black}{where \textbf{M} is a mobility matrix that depends on the diffusion coefficients of the Fokker-Planck dynamics\footnote{
\textcolor{black}{In general, there is freedom in setting the mobility matrix, and this freedom, along with the dependence on the Fokker-Planck dynamics, is discussed
in Section \ref{sec:entropy}.}}
\cite{PhysRevE.85.051113}.}

It is now evident that the connection with Wasserstein distance cannot be naturally restored unless one considers a second quantity, reflecting the effect of $\mathbf{J}^{\rm rev}$. Namely, while $\Sigma$ arises from currents that generate entropy, a second {\black action integral} $\Upsilon$, analogously defined as
$$
\Upsilon :=\int_{0}^{\tau}\int \cfrac{\|\mathbf{J}^{\text{rev}}\|_{\mathbf{M}}^2}{\rho}~\text{d}x\text{d}v\text{d}t,
$$  
arises from currents that do not. {\color{black} Thus, the purpose of this work is to explore the extent to which the intimate link between irreversibility, Wasserstein length, and speed limits for overdamped dynamics carries over to underdamped and to more general stochastic thermodynamic systems.

To this end, in the present paper, we develop the above formalism and explore time-symmetry properties of probability currents for general dynamics, to derive thermodynamic bounds for Langevin systems. In complete analogy with the overdamped regime, we utilize the fluid dynamic framework of Wasserstein geometry albeit in ``phase space''~\cite{Benamou2000ACF}.

In this broader context, we show that both action integrals, $\Sigma$ and $\Upsilon$, contribute to limit the speed of thermodynamic transitions via inequalities such as
}
\begin{subequations}
\begin{align}\label{eq:SU}
\Sigma + \Upsilon \geq \frac{1}{2\tau}\mathcal{W}_{2,\mathbf{M}}(\rho_0,\rho_\tau)^2,
\end{align}
{\black where $\mathcal{W}_{2,\mathbf{M}}(\rho_0,\rho_\tau)$ is an  $\mathbf{M}$-weighted Wasserstein distance (see eq.~\eqref{eq:BBM})}. Similar to \eqref{ODb}, \eqref{eq:SU} implies the speed limit
\begin{align}
   \lim_{\tau\to 0^+}\frac{\mathcal W_{2,\mathbf M}(\rho_t,\rho_{t+\tau})}{\tau}\leq \sqrt{2(\sigma_t +y_t)},  
\end{align}
where 
\begin{align*}
    \Sigma = \int_0^{\tau}\sigma_t\text{d}t~~\text{  and  }~~\Upsilon= \int_0^{\tau}y_t\text{d}t.
\end{align*}
A collection of such thermodynamic bounds are developed and specialized to underdamped Brownian particles subject to arbitrary forcing. {\black The paper concludes with two examples that highlight applications of the framework.}

\end{subequations}

\section{Preliminaries}

\subsection{On Optimal Mass Transport}
Let $\mathcal{P}_2(\mathbb{R}^n)$ denote the space of probability distributions with finite second-order moments on $\mathbb{R}^n$. Then,
\begin{equation}\nonumber
    {\mathcal{W}}_2(\rho_0,\rho_\tau) := \sqrt{\inf_{\pi \in  \Pi(\rho_0,\rho_\tau)} \int_{\mathbb R^n\times \mathbb R^n} \|\x-\y\|^2 \pi(\x,\y) \text{d}\x \text{d}\y} ,
\end{equation}
 defines a metric on $\mathcal{P}_2(\mathbb{R}^n)$ that is referred to as the $L^2$-Wasserstein metric. 
 In this definition, $\rho_0,\rho_\tau\in \mathcal{P}_2(\mathbb{R}^n)$ and  $\Pi(\rho_0,\rho_\tau)$ denotes the set of 
probability distributions on $\mathbb R^n\times \mathbb R^n$ having $\rho_0$ and $\rho_\tau$ as marginals.

The Wasserstein metric, originally interpreted as an optimal mass transfer cost, can be equivalently defined in the framework of continuum mechanics via the so-called Benamou-Brenier formulation \cite{Benamou2000ACF}. Specifically, for any time-interval $[0,~\tau]$, the square of the metric can be expressed as the minimizer of an action integral,
\begin{subequations}
{\black \begin{align}\label{eq:BB}
\!\!\!\!\!\!\mathcal{W}_2(\rho_0,\rho_\tau) &:= \sqrt{\inf_{\rho,u}\tau\int_{0}^\tau\!\!\int_{\mathbb{R}^n}\rho(t,\x)\|u(t,\x)\|^2\text{d}\x\text{d}t},
\end{align}}where minimization is {\black carried out} over time-dependent densities $\rho(t,\x)$ and velocity fields $u(t,\x)\in\mathbb{R}^n$ that together obey the continuity equation $$\partial_t\rho = -\nabla\cdot(\rho u),$$ for $t\in [0,~\tau]$, along with the marginal density constraints 
\begin{align*}
    \rho(0,\cdot)=\rho_0~~\text{ and }~~\rho(\tau,\cdot)=\rho_{\tau}.
\end{align*}
A \textit{weighted} Wasserstein metric can be defined in a similar manner, by suitably scaling
the Euclidean norm $\|u\|^2_{\mathbf{M}}:=u^T\mathbf{M}u$ with a positive-definite matrix $\mathbf{M}$, as
{\black \begin{align}\label{eq:BBM}
\!\!\!\mathcal{W}_{2,\mathbf{M}}(\rho_0,\rho_\tau) := \sqrt{\inf_{\rho,u}\tau\int_{0}^\tau\!\!\int_{\mathbb{R}^n}\rho(t,\x)\|u(t,\x)\|_{\mathbf{M}}^2\text{d}\x\text{d}t}.
\end{align}}
\end{subequations}It turns out that $\mathcal{W}_{2,\mathbf{M}}$ can be expressed directly in terms of $\mathcal{W}_{2}$, warping first the space of densities by $\mathbf{M}$ \cite{miangolarra2023minimal}, {\black i.e.
$$
\mathcal{W}_{2,\mathbf{M}}(\rho_0,\rho_\tau) = \mathcal{W}_{2}(\phi\sharp\rho_0,\phi\sharp\rho_{\tau}),
$$
with $\phi\sharp\rho$ being the {\em push-forward} of $\rho$ via the linear map $\phi\;:\;\x \mapsto \mathbf{M}^{1/2}\x$, that corresponds to a change of variables as dictated by the map $\phi$.}

\subsection{Stochastic Model} \label{sec:Model}
A prototypical example of Langevin dynamics with reversible and irreversible probability currents models the motion of Brownian particles at position $x(t)\in\mathbb R$ and velocity $v(t)\in\mathbb R$, and is given by
\begin{align*}
   \text{d}x &= v\text{d}t,\\
   m\text{d}v &= F(t,x,v)\text{d}t -\gamma v\text{d}t +\sqrt{2\gamma k_BT}\text{d}B_t,
\end{align*}
where $F$ is an arbitrary force and $B_t\in\mathbb{R}$ denotes standard Brownian motion. Throughout, $\gamma$ denotes the scalar friction coefficient, $k_B$ the Boltzmann constant, and $T$ the bath temperature. The corresponding probability density $\rho(t,x,v)$ obeys the Fokker-Planck equation
{\black \begin{align}\label{eq:FP2}
    \partial_t\rho & = -\nabla\cdot\mathbf{J},
\end{align}}where the components of the probability current $\mathbf{J}$ are 
\begin{subequations}\label{eq:currents-under}
    \begin{align}
    J_x &= v\rho,\\
    J_v &= \cfrac{1}{m}\left(F-\gamma v-\cfrac{\gamma k_B T}{m}\nabla_v\log\rho\right)\rho.
\end{align}
\end{subequations}
{\black Alternatively,  \eqref{eq:FP2} can be written as $\partial_t\rho=\mathcal{L}_t\rho$ with the (forward Kolmogorov) operator $ \mathcal{L}_t$ satisfying 
\[
\mathcal{L}_t\rho= -\nabla\cdot\mathbf{J}.
\]
}As alluded to in the introduction, the driving current $\mathbf{J}$ can be written as a sum of reversible and irreversible components denoted by $\mathbf{J}^{\text{rev}}$ and $\mathbf{J}^{\text{irr}}$, respectively. The irreversible current is then used to define the entropy production, which is seen as a measure of time-reversal symmetry-breaking of the dynamics \cite{PhysRevE.85.051113}. 

In general, and to that end, we first identify the parity of the variables and parameters in the system by classifying them as \textit{even} or \textit{odd} based on their sign change under the time reversal operation. In our case, $x$ is even and $v$ is odd. Then, we flip the sign of all odd variables and parameters in 
\begin{align*}
    \mathcal{L}_t= -v\nabla_x -\cfrac{1}{m}\nabla_v\left[\left(F-\gamma v\right)\cdot\,\right]+\cfrac{\gamma k_BT}{m^2}\Delta_v,
\end{align*}
to get the conjugate operator
\begin{align*}
    \mathcal{L}^{\dag}_t= \phantom{+}v\nabla_x +\cfrac{1}{m}\nabla_v \left[\left(F^{\dag}+\gamma v\right)\cdot\,\right]+\cfrac{\gamma k_BT}{m^2}\Delta_v.
\end{align*}
The notation $F^\dag$ implies that all odd variables and parameters in $F$ must also be flipped\footnote{For example, in some instances, when a magnetic field is present, its orientation is flipped under time reversal.}. By construction, the kinematic involution $\dag$ transforms $\mathcal{L}_t$ into an operator that corresponds to the dynamics responsible for generating the time-reversed paths in \textit{phase space}. By decomposing $\mathcal{L}$ into its odd and even parts with respect to $\dag$, we obtain the reversible {\black (odd)} and irreversible {\black (even)} evolution operators
$$
   \mathcal{L}_t^{\text{rev}}:= \cfrac{\mathcal{L}_t-\mathcal{L}_t^{\dag}}{2}~~~\text{and}~~~\mathcal{L}_t^{\text{irr}}:=\cfrac{\mathcal{L}_t+\mathcal{L}_t^{\dag}}{2},
$$
{\black
that reflect a time-symmetry, since 
\[
\left(\mathcal{L}_t^{\text{rev}}\right)^{\dag} = -\mathcal{L}_t^{\text{rev}}\text{ and }\left(\mathcal{L}_t^{\text{irr}}\right)^{\dag} = \mathcal{L}_t^{\text{irr}}.
\]}
Finally, the corresponding currents are identified from  
\begin{align*}
\mathcal{L}^{\text{rev}}_t\rho 
 = -\nabla\cdot\mathbf{J}^{\text{rev}}~~\text{and}~~\mathcal{L}^{\text{irr}}_t\rho 
 = -\nabla\cdot\mathbf{J}^{\text{irr}},
 \end{align*}
so that, in the present case,
\begin{subequations}
 \begin{align}
 \label{eq:Jrev}
 \mathbf{J}^{\text{rev}} &=\begin{pmatrix}
        v\\
        \cfrac{1}{m}F^{\text{rev}}
    \end{pmatrix}\rho,\\
    \label{eq:Jirr}
    \mathbf{J}^{\text{irr}} &=\begin{pmatrix}
        0\\   
        \cfrac{1}{m}F^{\text{irr}}-\cfrac{\gamma}{m}(v+\cfrac{k_BT}{m}\nabla_v\log\rho)
    \end{pmatrix}\rho,
\end{align}
\end{subequations}
where 
\begin{align*}
    F^{\text{rev}} = \cfrac{F+F^{\dag}}{2}~\text{ 
 and  }~F^{\text{irr}} = \cfrac{F-F^{\dag}}{2}.
\end{align*}
In general, the Fokker-Planck equation can be written as
\begin{align}
    \partial_t\rho = -\nabla\cdot\mathbf{J}^{\text{rev}} -\nabla\cdot\mathbf{J}^{\text{irr}}.\label{general}
\end{align} The current decomposition can also be obtained by defining an involution on $\mathbf{J}$ directly. This yields a conjugate current that satisfies $$\mathcal{L}^{\dag}\rho = -\nabla\cdot\mathbf{J}^{\dag},$$ leading to 
\begin{align*}
    \mathbf{J}^{\text{rev}} =\cfrac{\mathbf{J} - \mathbf{J}^{\dag}}{2}~~\text{  and  }~~ \mathbf{J}^{\text{irr}} = \cfrac{\mathbf{J} + \mathbf{J}^{\dag}}{2}.
\end{align*}
 The irreversible current $\mathbf{J}^{\text{irr}}$ is the only party responsible for the total entropy produced.

\subsection{Entropy production}
\label{sec:entropy}
During a thermodynamic transition, the total entropy production $\Sigma$ can be decomposed into 
$$\Sigma = \Sigma_{\text{sys}}+\Sigma_{\text{env}},$$ where $\Sigma_{\text{sys}}$ is the entropy produced in the system and $\Sigma_{\text{env}}$ is the entropy produced in the environment. The term $\Sigma_{\text{sys}}$ is defined to be the Shannon entropy difference between the initial and final states, and can also be written as 
\begin{align}\nonumber
    \Sigma_{\text{sys}} &=-k_B\int_{0}^{\tau}\int \partial_t\rho\log\rho~\text{d}x\text{d}v\text{d}t\\&=k_B\int_{0}^{\tau}\int \nabla\cdot \left(\mathbf{J}^{\text{rev}}+\mathbf{J}^{\text{irr}}\right)\log\rho~\text{d}x\text{d}v\text{d}t.
    \label{eq:Sigmasys}
\end{align}
    The second term can be further decomposed into  $$\Sigma_{\text{env}} = \Sigma_{\text{res}} + \Sigma_{\text{pu}},$$where $\Sigma_{\text{res}}$ is the entropy production mediated by the heat 
    dissipated into the reservoir \cite[Eq.\ (4.3)]{sekimoto2010stochastic}
\begin{align}\nonumber 
\Sigma_{\text{res}} 
&
= \cfrac{\gamma}{T}\int_0^{\tau}\hspace*{-6pt}\int \hspace*{-3pt}\left(v+\cfrac{k_BT}{m}\nabla_v\log\rho\right)\!v\rho\text{d}x\text{d}v\text{d}t\\
&= \cfrac{\gamma}{T}\int_{0}^{\tau}\langle v^2\rangle\text{d}t-\frac{\gamma k_B}{m}\tau,\label{eulg}
\end{align}
with the second equality following via integration by parts,
and $\Sigma_{\text{pu}}$ is the entropy ``pumped" into the environment by external forcing \cite{PhysRevE.85.051113}, \cite[Eq.\ (26)]{kwon2016unconventional},
\begin{align*}
 \Sigma_{\text{pu}} = &\int_0^{\tau}\int \left(\cfrac{(F^{\text{irr}})^2 - 2\gamma vF^{\text{irr}}}{\gamma T} +\cfrac{k_B}{m}\nabla_vF^{\dag}\right)\rho\text{d}x\text{d}v \text{d}t.
  \end{align*}  
The term $\Sigma_{\text{pu}}$ is typically present only when $F$ depends on $v$. More compactly, $\Sigma_{\text{env}}$ can be written in terms of the irreversible and reversible current components as 
\begin{align*}
    \Sigma_{\text{env}} &= \int_{0}^{\tau}\int\cfrac{\|\mathbf{J}^{\text{irr}}\|^2_{\mathbf{M}}}{\rho} ~\text{d}x\text{d}v\text{d}t\\
    &-k_B\int_0^{\tau}\int \nabla\cdot \left(\mathbf{J}^{\text{rev}}+\mathbf{J}^{\text{irr}}\right)\log\rho~\text{d}x\text{d}v\text{d}t,
\end{align*}
and as a result, the total entropy produced becomes
\begin{align*}
\Sigma =\int_{0}^{\tau}\int\cfrac{\|\mathbf{J}^{\text{irr}}\|^2_{\mathbf{M}}}{\rho} ~\text{d}x\text{d}v\text{d}t,
\end{align*}
where
\begin{equation}\label{eq:M}
\mathbf{M} = \frac{1}{\gamma T}\begin{bmatrix}\gamma^2 & 0\\0 & m^2
\end{bmatrix}
\end{equation}
is a temperature-scaled mobility matrix.
\textcolor{black}{In general, the Fokker-Planck dynamics $\partial_t\rho= -\sum_i\partial_{x_i}(A_i\rho)-\sum_i\partial^2_{x_i}(D_i\rho)$, dictate that, whenever $D_i\neq 0$, the corresponding diagonal entry of $\mathbf{M}$ is  $k_B/D_i$ \cite{PhysRevE.85.051113}. For this reason, the 2-2 entry in \eqref{eq:M} is specified, whereas the 1-1 entry can be selected as an arbitrary scaling factor with the appropriate units.}
   {\color{black}
   This freedom
    is explored in Section \ref{sec:coarsegraining}, while here we make the natural choice and select  $\mathbf M$ so that the ratio between the 1-1 and the 2-2 entries is the square of the characteristic time-scale $m/\gamma$.}

\section{Speed limits}

{\color{black} In this section we develop speed limits for thermodynamic transitions based on the Benamou-Brenier formulation of optimal mass transport. We begin in Section III.A by discussing the setting of general Langevin dynamics. In Section III.B, we further explore the consequences of the Benamou-Brenier formulation
by specializing to thermodynamic transitions in phase space for underdamped Brownian particles with arbitrary forcing.
}
\subsection{General Langevin dynamics}

For a finite time transition from $\rho_0$ to $\rho_\tau$, two quantities that characterize the effect of reversible and irreversible currents can be defined, specifically, the {\black{\em irreversible action}} (entropy production)
\begin{align*}
    \Sigma &= \int_{0}^{\tau}\int \cfrac{\|\mathbf{J}^{\text{irr}}\|_{\mathbf{M}}^2}{\rho}~\text{d}x\text{d}v\text{d}t,
\end{align*}
and the {\black{\em reversible action}}
\begin{align*}
    \Upsilon &= \int_{0}^{\tau}\int \cfrac{\|\mathbf{J}^{\text{rev}}\|_{\mathbf{M}}^2}{\rho}~\text{d}x\text{d}v\text{d}t.
    \end{align*}
 From \eqref{eq:BBM}, with $\x=(x,v)$ and $\mathbf J=\rho(t,\x)u(t,\x)$,
\begin{align*}
\tau\int_{0}^{\tau}\int\cfrac{\|\mathbf{J}\|_{\mathbf{M}}^2}{\rho}~\text{d}x\text{d}v\text{d}t \geq \mathcal{W}_{2,\mathbf{M}}(\rho_0,\rho_{\tau})^2.
\end{align*}
Substituting $\mathbf{J} =  \mathbf{J}^{\text{rev}} +\mathbf{J}^{\text{irr}}$ gives
 \begin{align}
    \mbox{\fbox{$\tau(\Sigma +\Phi+\Upsilon) \geq \mathcal{W}_{2,\mathbf{M}}(\rho_0,\rho_\tau)^2$}}\label{ASL}
\end{align}
where 
\begin{align}\label{eq:crossterm}
\Phi = 2\int_{0}^{\tau}\int\cfrac{(\mathbf{J}^{\text{rev}}\cdot\mathbf{J}^{\text{irr}})_{\mathbf{M}}}{\rho}~\text{d}x\text{d}v\text{d}t
\end{align}
{\black is a {\em cross-action}  integral, and $(\;\cdot\; )_{\mathbf M}$ denotes the inner-product with respect to $\mathbf M$}.
Inequality \eqref{ASL} is one of the main points of this contribution. Although rudimentary, it applies to general Langevin dynamics described by \eqref{eq:FP2} with a current that admits a decomposition into reversible and irreversible components\footnote{In the special case of overdamped dynamics when there are no odd variables/parameters, $\Upsilon = \Phi=0$ and \eqref{OD} follows for $\mathbf{M} = \gamma/T$. Likewise, for a deterministic system with reversible dynamics, $\tau\Upsilon \geq \mathcal{W}^2_{2,\mathbf{M}}(\rho_0,\rho_\tau)$.}. Furthermore, it implies the Wasserstein speed limit
\begin{align}\label{eq:rates}
\lim_{\tau\to 0^+}\cfrac{\mathcal{W}_{2,\mathbf{M}}(\rho_t,\rho_{t+\tau})}{\tau}\leq \sqrt{\sigma_t+\phi_t+y_t},
\end{align}
where 
\begin{align*}
    \sigma_t=\dot\Sigma,~~y_t=\dot\Upsilon,\text{  and  }\phi_t=\dot\Phi.
\end{align*}
In light of the fact that $|\phi_t|\leq \sigma_t+y_t$, \eqref{ASL} implies \eqref{eq:SU}.
The essence in \eqref{eq:rates} is that the speed of the Fokker-Planck flow in the Wasserstein metric, for a fairly general class of Langevin dynamics, is bounded by energetic costs of reversible and irreversible rates in this precise way. 

\textcolor{black}{
In the overdamped regime ($\mathbf{J}^{\text{rev}} = 0$) there is always an optimal choice for the force that saturates inequality \eqref{ASL}, and equivalently \eqref{OD}. However, this is not the case for more general dynamics. 
As discussed in \cite{PhysRevE.93.042112}, in the underdamped regime the ability to specify directly the entries of the probability current $\mathbf J$ is limited, since the relation between position and velocity is dictated by Newton’s equations of motion (see \eqref{eq:currents-under}). Indeed, the $J_x$ component of $\mathbf{J}$ can only be influenced by the applied force $F$ indirectly, by way of the system dynamics. In this sense, we do not have full control authority \cite{bechhoefer}, which is the ability to specify the components of $\mathbf{J}$ directly via a suitable selection of forces. Thus, there may not exist values for $F$ to saturate \eqref{ASL}, and the bound in \eqref{ASL} need not be tight. In what follows, \eqref{ASL} is specialized to the general underdamped dynamics introduced in \ref{sec:Model}, and corresponding thermodynamic inequalities are derived.}
\subsection{Underdamped dynamics}

Hereafter, we specialize \eqref{ASL} to the case of underdamped dynamics where\footnote{\black Throughout, the notation $\langle\;\rangle$ denotes phase-space expectation, integrating with respect to $\rho$.}   
\begin{align*}
    \Sigma =\cfrac{\gamma}{ T}\int_{0}^{\tau}\bigg\langle\bigg\|\cfrac{F_{\text{irr}}}{\gamma}-v-\cfrac{k_BT}{m}\nabla_v\log\rho\bigg\|^2\bigg\rangle \text{d}t,
\end{align*}
and 
\begin{align*}
\Upsilon = \cfrac{1}{\gamma T}\int_{0}^{\tau}\langle\|F^{\text{rev}}\|^2\rangle \text{d}t + \cfrac{\gamma}{T}\int_{0}^{\tau}\langle v^2\rangle \text{d}t,
\end{align*}
{\black
which gives a lower bound on the integral-square of externally applied forces, namely,
 \begin{align}
\mbox{\fbox{$\displaystyle \int_0^\tau \langle F^2\rangle \text{d}t   \geq
\int_0^{\tau}\langle\|\cfrac{\gamma k_BT}{m}\nabla_v\log\rho\|^2\rangle\text{d}t+\Theta$.}}\label{Gen}
 \end{align}
 The first term {\color{black}on the right-hand side} is a Fisher information functional, while the second
 \[
 \Theta = \frac{1}{\tau}\mathcal{W}^2_{2,\mathbf{N}}(\rho_0,\rho_{\tau}) + 2\gamma\left(\Delta  E_{\text{Kin}} - T\Sigma_{\text{sys}}-\frac{\gamma k_BT\tau}{m}\right),
 \]
 depends on boundary conditions. Specifically,
  $\Delta E_{\text{Kin}}$ is the change in the average kinetic energy of the particle between the endpoints and $\mathbf{N} = \text{diag}[\gamma^2,~m^2]$.
  The integral-square of forces can be interpreted as the ``control effort'' required to drive a Langevin system between two states in a fixed time $\tau$. 

  The bound applies with no restriction on the type of external forces $F(t,x,v)$, and its tightness can be studied with tools of minimum energy control \cite{chen2015optimal,chen2016relation,bechhoefer}.
  The proof of \eqref{Gen} is provided in Section \ref{sec:appendixA} of the Supplementary Material.}

  \subsubsection{Coarse graining}
\label{sec:coarsegraining}
 As noted earlier, there is a freedom in selecting $\mathbf M$ in \eqref{ASL}, which impacts both sides of the inequality.
Specifically, we obtain
the family of bounds
  \begin{align}
    \tau \geq \cfrac{\mathcal{W}_{2,\mathbf{M}_{\alpha}}(\rho_0,\rho_\tau)^2}{\Upsilon_{\alpha} +\alpha_v\left(\Phi+\Sigma\right)},
    \label{ASLG}
\end{align}
where 
\[
\mathbf{M}_{\alpha} := \frac{1}{\gamma T}\begin{bmatrix}
    \alpha_x\gamma^2&0\\
    0&\alpha_v m^2
\end{bmatrix},
\]
with $\alpha_x$,
$\alpha_v >0$ dimensionless parameters and 
\begin{align*}
     \Upsilon_{\alpha} = \int_{0}^{\tau}\int\cfrac{\|\mathbf{J}^{\text{rev}}\|_{\mathbf{M}_{\alpha}}^2}{\rho}~\text{d}x\text{d}v\text{d}t.
\end{align*}
{\black The choice of $\alpha_x/\alpha_v$ skews the metric, giving different importance to $\Upsilon_{\alpha}$ and $ \Phi+\Sigma$, and can be exploited to provide different bounds, as explained below.

In light of the fact that
\[
\mathcal{W}_{2,\mathbf{M}_{\alpha}}(\rho_0,\rho_\tau)^2
\geq
\frac{\alpha_x \gamma}{ T}\mathcal{W}_{2}(\rho_0^x,\rho_\tau^x)^2
+
\frac{\alpha_v m^2}{\gamma T}\mathcal{W}_{2}(\rho_0^v,\rho_\tau^v)^2,
\]
one can readily obtain speed limits for coarse-grained dynamics in each of the two degrees of freedom,
focusing on the corresponding marginal densities
\begin{align*}
    \rho^x=\int\rho dv\text{  and  }\rho^v=\int\rho dx,
\end{align*}
at the two end-points in time.
To this end, from \eqref{ASLG},
\begin{align}
\tau\geq\cfrac{\alpha_x\gamma^2\mathcal{W}_{2}(\rho^x_0,\rho^x_\tau)^2 + \alpha_vm^2\mathcal{W}_{2}(\rho^v_0,\rho^v_\tau)^2}{\displaystyle\int_{0}^{\tau}\int\cfrac{\alpha_x\gamma^2\|J_x\|^2+\alpha_vm^2\|J_v\|^2}{\rho}~\text{d}x\text{d}v\text{d}t},\label{coarsegrained}
\end{align}
and therefore, bounds on $\tau$ in the two cases follow by taking the limit of the right-hand side as  $\alpha_x\to 0$, and as $\alpha_v\to 0$, respectively.

It turns out that tighter bounds} can be obtained by employing \eqref{eq:BB} on the coarse-grained evolution equations
     \begin{align}
    \partial_t\rho^x = -\nabla_x\bar{J}_x~~\text{   and   }~~\partial_t\rho^v = -\nabla_v\bar{J}_v,\label{cfp}
\end{align}
where \begin{align*}
    \bar{J}_x = \int J_x \text{d}v~~\text{  and  }~~\bar{J}_v = \int J_v \text{d}x.
\end{align*}
{\black Comparison of the two sets of bounds, obtained via \eqref{coarsegrained} and \eqref{cfp}, respectively,} using the Cauchy-Schwarz inequality, is as follows:
\begin{subequations}
\begin{align}
\mbox{\fbox{$\displaystyle\tau\geq\cfrac{\mathcal{W}_2(\rho^x_0,\rho^x_\tau)^2}{\displaystyle \int_0^{\tau}\int\cfrac{\|\bar{J}_x\|^2}{\rho^x}\text{d}x\text{d}t}\geq\cfrac{\mathcal{W}_2(\rho^x_0,\rho^x_\tau)^2}{\displaystyle \int_0^{\tau}\int\cfrac{\|J_x\|^2}{\rho}\text{d}x\text{d}v\text{d}t}$}}\label{ex}\end{align}
and
\begin{align}
\mbox{\fbox{$ \displaystyle \tau\geq\cfrac{\mathcal{W}_2(\rho^v_0,\rho^v_\tau)^2}{\displaystyle \int_0^{\tau}\int\cfrac{\|\bar{J}_v\|^2}{\rho^v} \text{d}v\text{d}t} \geq  \cfrac{\mathcal{W}_2(\rho^v_0,\rho^v_\tau)^2}{\displaystyle \int_0^{\tau}\int\cfrac{\|J_v\|^2}{\rho} \text{d}x\text{d}v\text{d}t}$}}\label{vee}
\end{align}    
\end{subequations}

Both \eqref{ex} and \eqref{vee} imply lower bounds on $\Sigma$. Namely, the right-most bound in \eqref{ex}, combined with \eqref{eulg}, yields 
\begin{align}
 \Sigma \geq \cfrac{\gamma}{ T\tau}\mathcal{W}_2(\rho^x_0,\rho^x_\tau)^2+\Sigma_{\text{sys}}+\Sigma_{\text{pu}}-\frac{\gamma k_B}{m}\tau,\label{khod}
\end{align}
and the right-most bound in \eqref{vee} can be rearranged into
    \begin{align}
     \Sigma &\geq \cfrac{m^2}{\gamma T\tau}\mathcal{W}_{2}(\rho_0^v,\rho_{\tau}^v)^2 -\cfrac{1}{\gamma T}\int_0^\tau\langle \|F^{\text{rev}}\|^2\rangle \text{d}t - \Phi.\label{khod2}
\end{align} 
 For special cases of $F$, the bounds in \eqref{khod} and \eqref{khod2} become independent of the transition path. Briefly, if $F^{\text{irr}} = 0$ and $F$ is independent of $v$, then $\Sigma_{\text{pu}}= 0$ and the right-hand side of \eqref{khod} contains only boundary terms. Likewise, if $F^{\text{rev}}=0$, then $\Phi = 0$ and the right-hand side of \eqref{khod2} contains only boundary terms.
 \subsubsection{Special cases for $F$}
 We highlight results obtained from \eqref{ASLG} for the special cases where $F^{\rm rev}=0$ and $F^{\rm irr}=0$, respectively.\\[-0.1in]
 
\textbf{1) $F^{\text{rev}} = 0$.} This corresponds to a Brownian particle subject exclusively to forces that are odd under time reversal ($F^{\dag} = - F)$. Examples of such cases are the controlled friction in molecular refrigerator systems and forces on charged particles due to an \textit{irreversible} magnetic field.\footnote{While for the most part we consider one-dimensional Brownian particles, the framework readily extends to higher dimensions, where magnetic fields apply as a source of forcing.} Since $\Phi=0$, \eqref{ASLG} reduces to
\begin{subequations}
\begin{align}
     \mbox{\fbox{$\Sigma \geq \cfrac{1}{\tau}\mathcal{W}_{2,\mathbf{M}_{\alpha}}(\rho_0,\rho_\tau)^2 -  \cfrac{\gamma \alpha_x}{T}\int_{0}^{\tau}\langle v^2\rangle \text{d}t$}}\label{tight}
\end{align}
for $\alpha_x>0$ and $\alpha_v=1$. On the other hand, \eqref{vee} gives 
\begin{align}
\mbox{\fbox{$\displaystyle\Sigma \geq \cfrac{m^2}{\gamma T}\int_0^{\tau}\int\cfrac{\|\bar{J}_v\|^2}{\rho^v} \text{d}v\text{d}t \geq \cfrac{m^2}{\gamma T\tau}\mathcal{W}_2(\rho^v_0,\rho^v_\tau)^2$}}\label{margv}
\end{align}
\end{subequations}
It is not clear which of \eqref{tight} and \eqref{margv} is tighter in general. Focusing on \eqref{margv},
the right-most bound contains no information on the spatial distribution of the particles. Moreover, the middle expression involving $\bar J_v$ can be interpreted as coarse-grained entropy production, obtained by only observing the dynamics of the particle's velocity. This implies the intuitive assertion that coarse-graining can at most reduce the observed dissipation~\cite{dechant2019thermodynamic}. One might think that \eqref{margv} is trivial since in the absence of position-dependent forcing, the velocity dynamics become an overdamped system in $v$, making the saturation of the bound possible. The result in \eqref{margv}, however, still holds even when position-dependent forcing is present, as long as it is \textit{odd} under time reversal.\\[-.1in] 

\textbf{2) $F^{\text{irr}} = 0$.} This corresponds to Brownian particles subject to a force $F$ that is even under time reversal ($F^\dag = F$). Examples include forces due to a position-dependent potential or to a \emph{reversible} magnetic field. Interestingly, in general, when $F^\dag = F$, $\Phi$ becomes
\begin{align}
    \Phi &= -2\left(\cfrac{\Delta E_{\text{Kin}}}{T} +\Sigma_{\text{env}}\right),\label{sphi}
\end{align}
which is a key to subsequent inequalities.
The proof of \eqref{sphi}  can be found at the beginning of Section \ref{sec:appendixA} in the Supplementary Material. 

The expression in \eqref{sphi} combined with \eqref{ASLG} yields an upper bound on the entropy production,
\begin{subequations}
 \begin{align}\label{eq:23a}
  \mbox{\fbox{$ \displaystyle\Sigma \leq \Upsilon_{\alpha} +B_{\alpha} $}}
 \end{align}
that holds for all $(\alpha_x,\alpha_v)$ (that without loss of generality can be normalized to $\alpha_v=1$ and $\alpha_x > 0$), and where {\black $$B_\alpha = 2\Sigma_{\text{sys}}-\frac{2}{T}\Delta E_{\text{Kin}}  -\frac{1}{\tau}\mathcal{W}_{2,\mathbf{M}_{\alpha}}(\rho_0,\rho_\tau)^2$$}depends only on boundary conditions. Bounding the squared Wasserstein distance in \eqref{eq:23a} by the sum of the coarse-grained marginals as in \eqref{coarsegrained}, and taking the limit $\alpha_x\to 0$, we obtain that
 \begin{align}\label{eq:23b}
      \mbox{\fbox{$ \displaystyle\Sigma\leq \cfrac{1}{\gamma T}\int_0^\tau\langle \|F^{\text{rev}}\|^2\rangle \text{d}t +B$}}
 \end{align} 
\end{subequations}
where {\black$$ B = 2\Sigma_{\text{sys}}-\frac{2}{T}\Delta E_{\text{Kin}}-\frac{m^2}{\gamma T\tau}\mathcal{W}_{2}(\rho_0^v,\rho_{\tau}^v)^2. $$}The same expression follows from \eqref{khod2} by substituting \eqref{sphi} and rearranging terms.
 The last expression coincides with an upper bound on entropy production rate derived recently in \cite{dechant2023upper} for {\em position-dependent forcing at steady-state}. Here we include both  \eqref{eq:23a} and \eqref{eq:23b} since it is not clear that one is tighter than the other.

 We conclude with two lower bounds, one on control effort and one on entropy production.
 The first,
\begin{align}
\mbox{\fbox{$\displaystyle\int_{0}^{\tau}\langle F^2\rangle \text{d}t\geq  \gamma T\Sigma_{\text{pu}} + \Gamma$}}\label{simil}
\end{align}
follows by substituting \eqref{sphi} into \eqref{ASL} (see detailed steps in Section \ref{sec:appendixB} of the Supplementary Material),
where {\black$$\Gamma = \frac{1}{\tau}\mathcal{W}_{2,\mathbf{N}}(\rho_0,\rho_\tau)^2 + \gamma \left(2\Delta E_{\text{Kin}} -T\Sigma_{\text{sys}} -\frac{\gamma k_BT\tau}{m}\right),$$} depends on boundary conditions.
The right-hand side of \eqref{simil} contains $\Sigma_{\text{pu}}$, which typically depends on the transition path. However, in many cases where $\nabla_v F^{\text{rev}}=0$, $ \Sigma_{\text{pu}} = 0$ and the right-hand side of \eqref{simil} contains only boundary terms.
The second,
\begin{align}
\mbox{\fbox{$\displaystyle\Sigma \geq \cfrac{\gamma^2}{\gamma T}\int_0^{\tau}\int\cfrac{\|\bar{J}_x\|^2}{\rho^x}\text{d}x\text{d}t \geq \cfrac{\gamma^2}{ \gamma T\tau}\mathcal{W}_2(\rho^x_0,\rho^x_\tau)^2$}}\label{margx}
\end{align}
is analogous to \eqref{margv} in that the bound contains no information on the distribution of particle velocities. This result was first derived in \cite{dechant2019thermodynamic} for the special case where $F^{\text{rev}}$ is independent of $v$, although it holds without this condition. As explained in \cite{dechant2019thermodynamic},  the bound in \eqref{margx} is the natural analog of the overdamped version in \eqref{OD}, though it is not tight. Once again, now, the middle expression involving $\bar J_x$ in \eqref{margx} can be interpreted as a coarse-grained entropy production obtained by only observing the dynamics of the spatial distribution of particles. Finally, if in the typical case where $F^{\text{rev}}$ is independent of $v$, $\Sigma_{\text{sys}}\geq \gamma k_B\tau/m$, \eqref{khod} becomes tighter than \eqref{margx}. Nevertheless, it is interesting to note the symmetry of the results in \eqref{margv} and \eqref{margx}, each of which holds for arbitrary $F$, as long as $F$ has a definite parity under time reversal.

     \begin{figure}[H]
    \centering
    \includegraphics[width = 8cm]{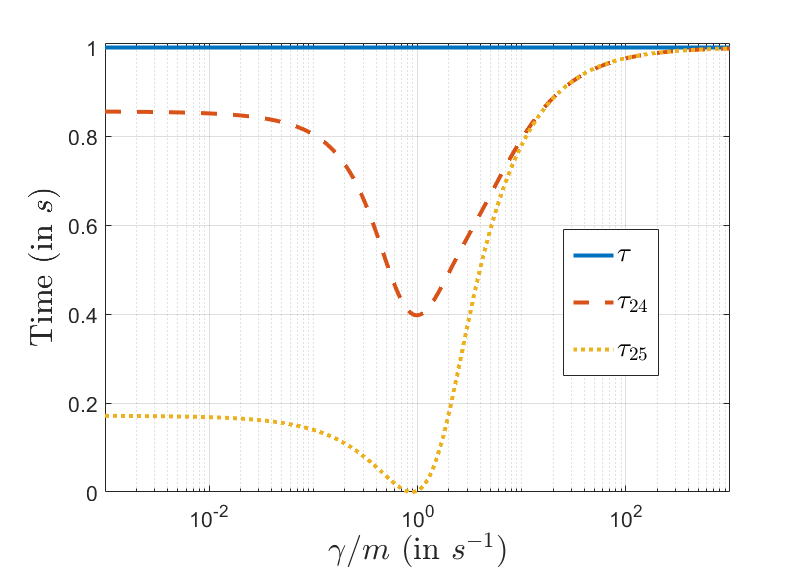}
    \caption{\color{black}Lower bounds $\tau_{24}$ (dashed) and $\tau_{25}$ (dotted) on the transition time $\tau$ (solid) for different values of $\gamma/m$ using the optimal protocol in the overdamped limit.}
    \label{fig:1}
\end{figure}
{\color{black}
\section{Examples}
\subsection{Quadratic time-varying potential}
 In this example, we compare the tightness of the speed limits obtained from inequalities \eqref{simil} and \eqref{margx} for different values of $\gamma/m$. To that end, consider a particle with mass $m = 10\,\rm [ng]$ confined in the quadratic time-varying potential 
\begin{align*}
    U(t,x) = \cfrac{1}{2}\,q(t)x^2,~t\in[0,\tau],
    \end{align*}
where $q(t)$ denotes the control protocol of the potential so that $F=-\nabla_xU = -q(t)x$.

In this case, $F^\dag = F$. By defining  
 the time-averaged control~effort 
   ($CE$) and the time-averaged entropy production ($EP$) over the transition as
\begin{align*}
    {\black
    CE}\equiv \cfrac{1}{\tau}\int_{0}^{\tau}\langle F^2\rangle \text{d}t~~\text{and}~~ EP\equiv \cfrac{\Sigma}{\tau},
\end{align*} inequalities \eqref{simil} and \eqref{margx} can be rearranged into
\begin{align*}
     a\tau^2+b\tau + c\geq 0\text{ and }\tau^2 \geq d,
\end{align*}
respectively, where 
 \begin{align*}
  a &=  CE+\frac{\gamma^2 k_BT}{m},~b =\gamma T\Sigma_{\text{sys}}-2\gamma\Delta E_{\text{Kin}},\\ c&=-\mathcal{W}_{2,\mathbf{N}}(\rho_0,\rho_\tau)^2,~ d = \frac{\gamma}{T(EP)}\mathcal{W}_2(\rho^x_0,\rho^x_\tau)^2.   
\end{align*}
Inequalities
 \eqref{simil} and \eqref{margx} yield the lower bounds $\tau_{24}$ and $\tau_{25}$ on the transition time, respectively, where
\begin{align*}
    \tau &\geq \tau_{\ref{simil}}\equiv\cfrac{-b + \sqrt{b^2 - 4ac}}{2a},\\
     \tau &\geq \tau_{\ref{margx}}\equiv\sqrt{d}.
\end{align*}

We compare the bounds $\tau_{24}$ and $\tau_{25}$ for different values of $\gamma/m$, as illustrated in Figure \ref{fig:1}. To this end, we consider the control protocol 
\begin{align*}
    q(t) =\cfrac{4k_BT}{(2-t)^2}+\cfrac{\gamma}{2-t}.
\end{align*}
This control law is the minimum-entropy protocol required to steer the system from an initial state $\rho_0^x \sim \mathcal{N}(0,I)$ to 
a final state $\rho_\tau^x \sim \mathcal{N}(0,0.5I)$ in $\tau = 1$ [sec] 
in the overdamped limit ($\gamma/m \gg {\color{black}1/}\tau$) \cite{8825523}. Here, we take $T = 295$K and $k_B = 1.38\times 10^{-23}$ J/K. 

Figure \ref{fig:1} shows that both bounds converge to $\tau=1$ as $\gamma/m$ increases.
This is expected since the control protocol has been chosen to be optimal in the overdamped regime where the full control authority over the probability current is regained\footnote{\color{black}
 The probability current in the overdamped limit $J_x=-\frac{1}{\gamma}\rho^x(\nabla_x U + k_BT\nabla_x \log \rho^x)$ can be fully specified by a suitable choice of $U$, cf.\ comment at the end of Section III.A.
}.
The bound $\tau_{24}$ appears less conservative than $\tau_{25}$ which may be attributed to utilizing finer information about the system.
An interesting behavior
of both bounds is observed \textcolor{black}{near} $\gamma/m=1$, where the transition between the low and the high friction regimes takes place.

\subsection{Noisy RLC circuit}
While the exposition of \eqref{ASL} has been carried out for the dynamics of an underdamped Brownian particle subject to a general force $F$, we emphasize that the framework is applicable to general Langevin dynamics with even and odd degrees of freedom as outlined in the introduction. We illustrate this by considering the noisy RLC circuit shown in Fig.\ \ref{RLC}. Specifically, we consider a time-varying inductor and a resistor in the network that is in contact with a heat bath of temperature $T$. 
\begin{figure}[h]
    \centering
\includegraphics[width=.65\linewidth]{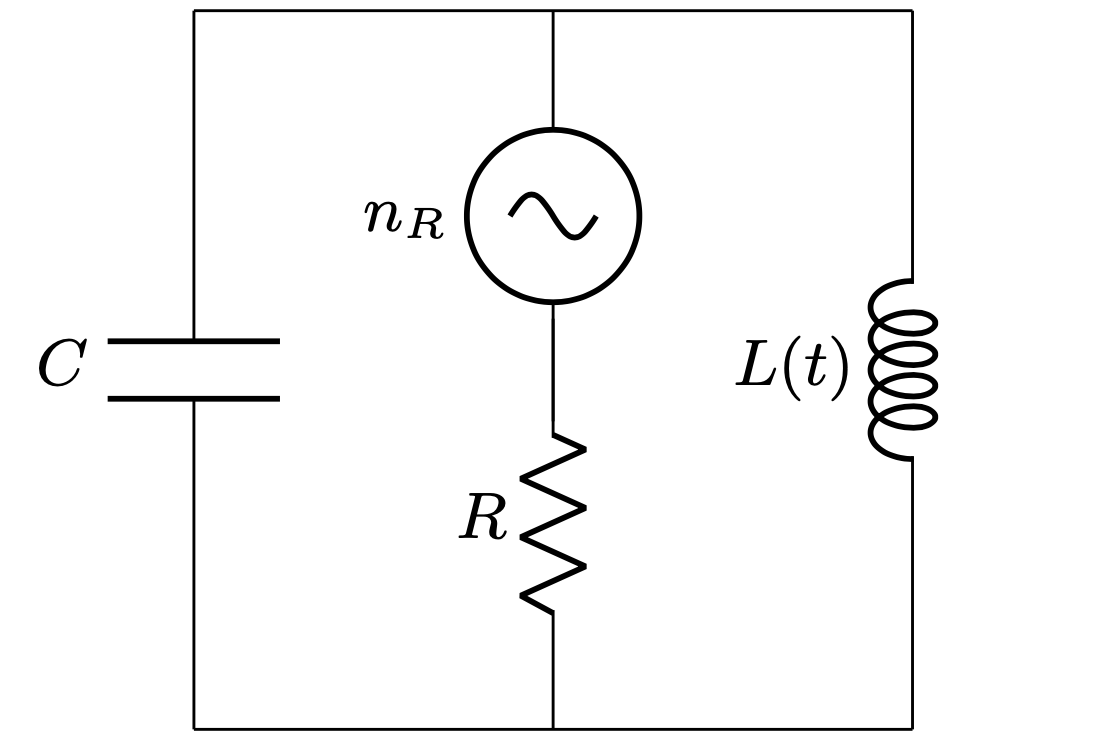}
    \caption{Noisy RLC Circuit with time-varying inductor.}
    \label{RLC}
\end{figure}

The charge $q$ stored at the capacitor together with the magnetic flux $\phi$ in the inductor follow second-order stochastic dynamics given by
\begin{align*}
    \text{d}\phi &= \frac{q}{C}\text{d}t,\\
    \text{d}q    &= \left(-\frac{q}{CR}-\frac{\phi}{L(t)}\right)\text{d}t+\sqrt{\frac{2k_BT}{R}}\text{d}B_t,
\end{align*}
where $C$ is the value of the capacitance, $L(t)$ the time-varying inductance and $R$ the resistance. The corresponding probability density $\rho(t,\phi,q)$ obeys the Fokker-Planck equation
\begin{align*}
    \partial_t\rho &= \mathcal{L}_t\rho = -\nabla\cdot\mathbf{J},
\end{align*}
where the components of the probability current $\mathbf{J}$ are 
\begin{align*}
    J_\phi &= \frac{q}{C}\rho,\\
    J_q &= \left(-\frac{q}{CR}-\frac{\phi}{L(t)}-\cfrac{ k_B T}{R}\nabla_q\log\rho\right)\rho.
\end{align*}
It was initially recognized by Casimir \cite{RevModPhys.17.343}, in his study of dynamic reversibility, that one should treat $q$ and $\phi$ as even and odd state variables respectively when comparing probability densities, much like the setting for underdamped Langevin dynamics. This idea was further validated in \cite{1084744} for nonlinear circuits. As a result, we obtain the following decomposition of the probability current 
\begin{subequations}
 \begin{align*}
 \mathbf{J}^{\text{rev}} &=\begin{pmatrix}
        q/C\\
       -\phi/L(t)
    \end{pmatrix}\rho,\\
    \mathbf{J}^{\text{irr}} &=\begin{pmatrix}
        0\\   
        -\cfrac{q}{CR}-\cfrac{k_BT}{R}\nabla_q\log\rho
    \end{pmatrix}\rho,
\end{align*}
\end{subequations}
which leads to the reversible and irreversible actions 
\begin{align*}
    \Sigma =\cfrac{1}{ RT}\int_{0}^{\tau}\bigg\langle\bigg\|\cfrac{q}{C}+k_BT\nabla_q\log\rho\bigg\|^2\bigg\rangle \text{d}t,
\end{align*}
and 
\begin{align*}
\Upsilon = \cfrac{1}{RTC^2}\int_{0}^{\tau}\langle q^2\rangle \text{d}t + \cfrac{R}{T}\int_{0}^{\tau}\bigg\langle \frac{\phi^2}{L(t)^2}\bigg\rangle \text{d}t,
\end{align*}
respectively, where  \begin{align*}
    \mathbf{M} = \begin{bmatrix}
        1/RT&0\\0&R/T
    \end{bmatrix}.
\end{align*}
Moreover, the cross-term is given by 
\begin{align*}
    \Phi &= \cfrac{2}{TC}\int_0^{\tau}\bigg\langle \cfrac{q\phi}{L(t)}\bigg\rangle\text{d}t.
\end{align*}
By noting that the entropies produced in the system and environment are respectively given by 
    \begin{align*}
    \Sigma_{\text{sys}} &=\cfrac{1}{RT}\int_{0}^{\tau}\bigg\langle \frac{qk_BT}{C}\nabla_q\log\rho+\bigg\|k_B T\nabla_q\log\rho\bigg\|^2 \bigg\rangle\text{d}t\\
    &=\cfrac{1}{RT}\int_{0}^{\tau}\bigg\langle \bigg\|k_B T\nabla_q\log\rho\bigg\|^2 \bigg\rangle\text{d}t - \cfrac{k_B\tau}{RC},\\
    \Sigma_{\text{env}}&=\cfrac{1}{RT}\int_{0}^{\tau}\bigg\langle \frac{qk_BT}{C}\nabla_q\log\rho+\left(\cfrac{q}{C}\right)^2\bigg\rangle\text{d}t\\
    &=\cfrac{1}{RTC^2}\int_{0}^{\tau}\langle q^2\rangle\text{d}t - \cfrac{k_B\tau}{RC},
\end{align*}
the cross-term can be expressed as 
\begin{align}
  \Phi = -2\left(\cfrac{\Delta E_{\text{Cap}}}{T}+\Sigma_{\text{env}}\right),\label{crs}
\end{align}
where $\Delta E_{\text{Cap}}$ is the change in the average electrical energy stored in the capacitor between $t = 0$ and $t = \tau$. Note the similarity between \eqref{crs} and \eqref{sphi}. Thus, substituting \eqref{crs} in \eqref{ASL} yields 
 \begin{align}
    \int_{0}^{\tau}\bigg\langle \frac{\phi^2}{L(t)^2}\bigg\rangle \text{d}t \geq \cfrac{1}{\tau} \mathcal{W}_{2,\mathbf{N}}(\rho_0,\rho_\tau)^2 +\cfrac{B}{R}~,\label{CEC}
\end{align}
where $\mathbf{N} = \text{diag}([1/R^2~~1])$ and \begin{align*}
    B = 2\Delta E_{\text{Cap}}- T\Sigma_{\text{sys}} -\cfrac{k_BT\tau}{RC}.
\end{align*}
The right-hand-side in \eqref{CEC} can be interpreted as a path-independent lower-bound on the minimum control effort required to steer a system from $\rho_0$ to $\rho_{\tau}$ through a time-varying protocol $L(t)$. Minimizing control effort relates to suppressing the thermal fluctuations of the magnetic flux.

This exemplifies an issue that arises in high-resolution instrumentation, where thermal fluctuations degrade the quality of measurements. In this context, feedback control has been used to mitigate the effect of thermal noise
\cite{PhysRevLett.103.010601,chen2015fast}.

\section{Concluding remarks}

The usage of optimal mass transport for quantifying universal speed limits for the evolution of physical systems on discrete spaces was pioneered in \cite{PhysRevLett.130.010402,van2023thermodynamic}. The framework in \cite{PhysRevLett.130.010402,van2023thermodynamic} utilized the Wasserstein $\mathcal W_1$ metric. In contrast, for continuous spaces where action integrals and control costs are typically quadratic, the $\mathcal W_2$  geometry  pursued herein appears natural \cite{dechant2019thermodynamic}.

{\black In the present paper we presented a universal bound \eqref{ASL} that is} valid for general Langevin systems, provided the Fokker-Planck equation admits a decomposition
$$
\partial_t\rho = -\nabla\cdot\mathbf{J} = -\nabla\cdot\mathbf{J}^\text{irr} -\nabla\cdot\mathbf{J}^\text{rev},
$$
with $\mathbf{J}^{\text{irr}}$ and $\mathbf{J}^{\text{rev}}$ {\black being the even and odd parts of $\mathbf{J}$ with respect to an involution $\dagger$ of the underlying dynamics}.
For overdamped Brownian particles where typically positional variables and (force) parameters are even, the bound in \eqref{ASL} reduces to the well-known bound on entropy production for overdamped systems \eqref{OD}, {\black namely,
\begin{align*}
    \tau\Sigma \geq \mathcal{W}_{2,\mathbf M}(\rho_0,\rho_\tau)^2.
\end{align*}
This inequality is characteristic of systems driven by a $\dag$-even (or irreversible) current; tightness of the inequality, which holds when sufficient control authority is available, allows expressing minimal irreversible action (entropy production) in terms of the Wasserstein length.
For systems driven by $\dag$-odd (or reversible) currents a mirror relation holds in that the bound \eqref{ASL} gives 
\[
\tau\Upsilon \geq \mathcal{W}_{2,\mathbf{M}}(\rho_0,\rho_\tau)^2,
\]
for the reversible action $\Upsilon$.

The purpose of this work has been to explore the extent to which the intimate connection between irreversibility, Wasserstein length, and speed limits, which is well established for overdamped dynamics, carries over to more general dynamics. The results that we presented suggest a more nuanced picture, where entropy production alone is not the main obstacle that curtails the speed of transition and where tighter bounds can be obtained by deeper analysis of reversible currents.
}

\bibliographystyle{unsrt}
\bibliography{references_LM18419W}

\newpage

\thispagestyle{empty}

\onecolumngrid

\section*{Supplementary Material}\label{sec:appendix}
\subsection{Proof of \eqref{Gen}} \label{sec:appendixA} For a general force $F$, we have 
\begin{align*}
    \Phi &= \cfrac{2}{T}\int_0^{\tau}\langle \cfrac{F^{\text{rev}}F^{\text{irr}}}{\gamma}-F^{\text{rev}}v-\cfrac{k_BT}{m}F^{\text{rev}}\nabla_v\log\rho\rangle\text{d}t= \Lambda_1 + \Lambda_2 + \Lambda_3,
\end{align*}
where 
\begin{align*}
    \Lambda_1 &= \cfrac{1}{\gamma T}\int_0^\tau\langle F^2-F^{\text{rev}^2}\rangle\text{d}t,\\
    \Lambda_2 &= -\cfrac{2}{T}\left(\Delta E_{\text{Kin}}+\Sigma_{\text{env}}T\right),\\
    \Lambda_3 &= \Sigma -\cfrac{\gamma}{T}\int_0^\tau\langle \|v+\cfrac{k_BT}{m}\nabla_v\log\rho\|^2\rangle\text{d}t.
\end{align*}
Notice that if $F^{\text{irr}}= 0$ then $\Lambda_1 = \Lambda_3 = 0$, and $
    \Phi = \Lambda_2 = -\cfrac{2}{T}\left(\Delta E_{\text{Kin}}+\Sigma_{\text{env}}T\right),
$ 
which is \eqref{sphi}. Moving on, the general expression for $\Phi$ combined with \eqref{ASL} yields
\begin{align*} \Upsilon+\Phi+\Sigma &=
    \cfrac{ \gamma}{T}\int_0^{\tau}\langle v^2\rangle \text{d}t+\cfrac{1}{\gamma T}\int_0^\tau\langle F^2\rangle\text{d}t-\cfrac{2}{T}\left(\Delta E_{\text{Kin}}-\Sigma_{\text{sys}}T\right)
    -\cfrac{\gamma}{T}\int_0^{\tau}\langle\|v+\cfrac{k_BT}{m}\nabla_v\log\rho\|^2\rangle\text{d}t\\& \geq \cfrac{1}{\tau}\mathcal{W}_{2,\mathbf{M}}(\rho_0,\rho_\tau)^2,
\end{align*}
so that 
\begin{align}
\cfrac{1}{\gamma T}\int_0^\tau\langle F^2\rangle\text{d}t -\cfrac{\gamma}{T}\int_0^{\tau}\langle\|\cfrac{k_BT}{m}\nabla_v\log\rho\|^2\rangle\text{d}t
\geq \cfrac{1}{\tau}\mathcal{W}_{2,\mathbf{M}}(\rho_0,\rho_\tau)^2+2\left(\Delta E_{\text{Kin}}/T -\Sigma_{\text{sys}}-\gamma k_B\tau/m\right),\label{final}
\end{align}
  which is \eqref{Gen}.
 \subsection{Proof of \eqref{simil}}\label{sec:appendixB}
 Using (\ref{eq:Jrev}-\ref{eq:Jirr}) and the fact that $F^{\text{irr}}= 0$, expand the right hand side of 
\eqref{eq:Sigmasys} 
 to obtain that $\Sigma_{\rm sys}$ equals
 \[
 \frac{k_B}{m}\int_0^\tau\!\!\!\int\log\rho\nabla_v\cdot (\rho F^{\rm rev}-\gamma \rho v- \frac{\gamma k_BT}{m}\rho\nabla_v \log\rho) \text d x\text d v\text d t.
 \]
 Use integration by parts and collect terms to obtain
 \begin{align*}
\Sigma_{\text{sys}} = \Sigma_{\text{pu}} +\cfrac{\gamma k_B}{m}\tau - \cfrac{\gamma}{T}\int_0^{\tau}\langle\|\cfrac{k_BT}{m}\nabla_v\log\rho\|^2\rangle\text{d}t,
 \end{align*}
since 
\[
\int_0^\tau\!\!\!\int v\nabla_v \rho \,\text d x\text d v\text d t=-\tau,
\] 
 by using integration by parts one more time.
Substituting
the ``Fisher information'' term
\[
\cfrac{\gamma}{T}\int_0^{\tau}\langle\|\cfrac{k_BT}{m}\nabla_v\log\rho\|^2\rangle\text{d}t
=
-\Sigma_{\text{sys}} + \Sigma_{\text{pu}} +\cfrac{\gamma k_B}{m}\tau
\]
in the left hand side of \eqref{final} yields \eqref{simil}.
\end{document}